\documentclass[11pt]{article}
\usepackage{epsf}
\usepackage{amssymb}
\usepackage{latexsym}
\usepackage{amsmath}

\def\Z{\Bbb Z}

\def\ra{\rightarrow}

\def\bsk{\bigskip}
\def\ds{\displaystyle}

\def\lan{\langle}
\def\ran{\rangle}
\newcommand{\ph}{\phantom{\ds{\frac{\frac{I}{I}}{\frac{I}{I}}}}}
\newtheorem{lemma}{Lemma}
\newtheorem{theorem}[lemma]{Theorem}
\begin{document}
\title{\large{\bf A Mourre estimate for a Schr\"odinger operator \\
                                        on a binary tree}}
\author{C. Allard and R. Froese \\ Department of Mathematics \\ 
               University of British-Columbia, Vancouver BC}
\date{July 8, 1998}
\maketitle
  \begin{quote}
   {\small \begin{center} {\bf Abstract} \end{center}
   \hspace*{0.4cm} {\sl Let $G$ be a binary tree with vertices $V$ and let 
   $H$ be a Schr\"odinger operator acting on 
   $\ell^{2}(V)$. A decomposition of the space 
   $\ell^{2}(V)$ into invariant subspaces is exhibited 
   yielding a conjugate operator $A$ for 
   use in the Mourre estimate. We show that for potentials $q$ 
   satisfying a first order difference decay condition, a Mourre 
   estimate for $H$ holds.}}
   \end{quote}

\vspace*{1cm}

\noindent {\bf Introduction}

\bsk
Let $G = (V,E)$ be a graph with vertices $V$ and edges $E$. The Laplace 
operator acts on functions defined on $V$. If $\phi:V \ra {\Bbb C}$ is 
such a function, then $\Delta \phi$ is the function defined by
$$
(\Delta \phi)(v) = \ds{\sum_{w:w-v}} (\phi(w) - \phi(v)),
$$
where $w-v$ means that $v$ and $w$ are connected by an edge. We are 
interested in the spectral \noindent theory of $-\Delta$ and 
perturbations $-\Delta + q$, acting in the Hilbert space $\ell^{2}(V)$ of 
square summable functions on $V$. This is the space of functions $\phi$ 
satisfying
$$
{\ds \sum_{v \in V}} |\phi(v)|^{2} < \infty
$$
with inner product given by
$$
\lan \phi, \psi \ran = \ds{\sum_{v \in V}} \overline{\phi}(v)\psi(v).
$$

Let $L$ denote the off-diagonal part of $\Delta$. Thus
$$
(L\phi)(v) = \ds{\sum_{w:w-v}}\phi(w).
$$

\noindent If $d(v)$ denotes the number of edges joined to the vertex $v$ 
then $$\Delta = L - d$$ where $d$ is the operator of multiplication 
by $d(v)$. The degree term $d$ can be included in the potential as a 
perturbation, hence $-\Delta + q = -L + d + q$ can be considered as a 
perturbation of $L$.
Both $\Delta$ and $L$ are symmetric on $\ell^{2}(V)$. When $d(v)$ is 
bounded, then both operators are also bounded operators, hence self-adjoint.

The goal of this paper is to prove a Mourre estimate and related bounds for 
the Schr\"odinger operator $-\Delta + q$ when the underlying graph is a 
binary tree. Here $q$ denotes multiplication by a potential function that 
tends to zero at infinity. For a binary tree, $d= 3 - d_{0}$, where 
$d_{0}$ is the potential with $d_{0}(v) = 1$ at the root of the 
tree and 0 otherwise. Hence $-\Delta + q = - L + 3 - d_{0} + q$ and 
the spectrum of $-\Delta + q$ is the same as that of $L - q + d_{0}$ 
up to a shift and a reflection about zero. In considering the Mourre 
estimate, the $d_{0}$ term can be absorbed in $q$ and the sign of the potential
changed, since $- q + d_{0}$ satisfies our decay assumptions 
whenever $q$ does. Hence we aim at obtaining a Mourre estimate for 
$L + q$.

The operator $L$ can be diagonalized explicitly. Its spectrum is absolutely 
continuous and equal
$\sigma(L)=\sigma_{\text{ac}}(L) = [-2\sqrt{2}, 2\sqrt{2}]$. This is 
also the essential spectrum of $L$. Since $q$ is a compact operator, 
perturbation by $q$ does not change the essential spectrum, and so 
$\sigma_{\text{ess}}(L + q) = [-2\sqrt{2}, 2\sqrt{2}]$.

We will define a self-adjoint conjugate operator $A$ such that, under 
appropriate conditions on the potential $q$ \par
(i) $[L+q,iA]$ is bounded \label{q1} \par
(ii) $[[L+q,iA],iA]$ is bounded \label {q2} \par
(iii) $L + q$ and $A$ satisfy a Mourre estimate at every point in 
      $(-2\sqrt{2}, 2\sqrt{2})$ \label{q3} \par
By definition (iii) means that for every 
$\lambda \in (-2\sqrt{2}, 2\sqrt{2})$, there exist an interval $I$ 
containing $\lambda$ such that
$$ E_{I}[L + q, iA]E_{I} \geq \alpha E^{2}_{I} + K$$
\noindent Here $E_{I} = E_{I}(L + q)$ denotes the (possibly smoothed) 
spectral projection corresponding to the interval $I$, $\alpha$ is a 
positive number, and $K$ is a compact operator. Precise statements can 
be found in Lemma \ref{l5}, Lemma \ref{l6}, Lemma \ref{l7} and Theorem 
\ref{t9} below.

The estimates (i) (ii) and (iii) together with the abstract Mourre theory
(see for example \cite{CFKS}), have the following consequences:\par
(1) Eigenvalues of $L + q$ not equal to $\pm 2 \sqrt{2}$ have finite 
    multiplicity and can only accumulate at $\pm 2 \sqrt{2}$.\par
(2) The operator $L + q$ has no singular continuous spectrum.\par
(3) Scattering for the pair $L$ and $L + q$ is asymptotically complete,
see \cite{A}. \par
Although we only treat the binary tree, the same method can be applied to 
related graphs, for example the Bethe Lattice or trees with $k$-fold 
branching. Schr\"odinger operators on the Bethe Lattice are of interest in 
solid state physics, where they serve as a model for tightly
bound electrons. Much effort has gone into studying operators with random 
potentials, and it is interesting to note that although the existence of 
dense point spectrum near the band edges has been proven in many situations,
the Bethe Lattice is the only model where it has been proved that for weak 
disorder, some absolutely continuous spectrum remains in the middle of the 
band \cite{K}. From the purely mathematical point of view, the Bethe
Lattice is the Cayley graph of a free group. It would be most 
interesting to be able to say something about the continuous spectrum of 
the Laplace operator on the Cayley graph for a finitely generated group that
is not free, and to relate properties of the spectrum to properties of the 
group. 

The subspace decomposition and conjugate operator used in this paper 
bear some similarity to the ones used in \cite{FH} in the case of 
exponentially large manifolds. However, the details are quite
different. In particular, the calculation of the 
matrix elements of $A$ and the method of estimating $[q,iA]$ have no 
analogue. These results first appeared in \cite{A}. 

\bsk

\noindent {\bf The operators $\Pi$ and $R$}

\bsk

In this section we will let $(V,E)$ be an arbitrary graph and introduce 
polar co-ordinates and some associated operators. Choose some $0 \in V$ to
be the origin. Define $|v|$ to be the distance in the graph from 0 to $v$. 
In other words, $|v|$ is the length of the shortest path in the graph 
joining 0 to $v$. Define $S_{r}$, the sphere of radius $r$, to be the set 
of all vertices with $|v| = r$. Then $V$ is a disjoint union
$$
V = \ds{ \bigcup_{r=0}^{\infty}} \ S_{r}
$$
and
$$
\ell^{2}(V) = \ds{\bigoplus_{r=0}^{\infty}} \ \ell^{2}(S_{r})
$$

In the case of the binary tree see figure below.

\begin{figure}[h]
\centerline{\epsfysize=5.5cm
\epsfbox{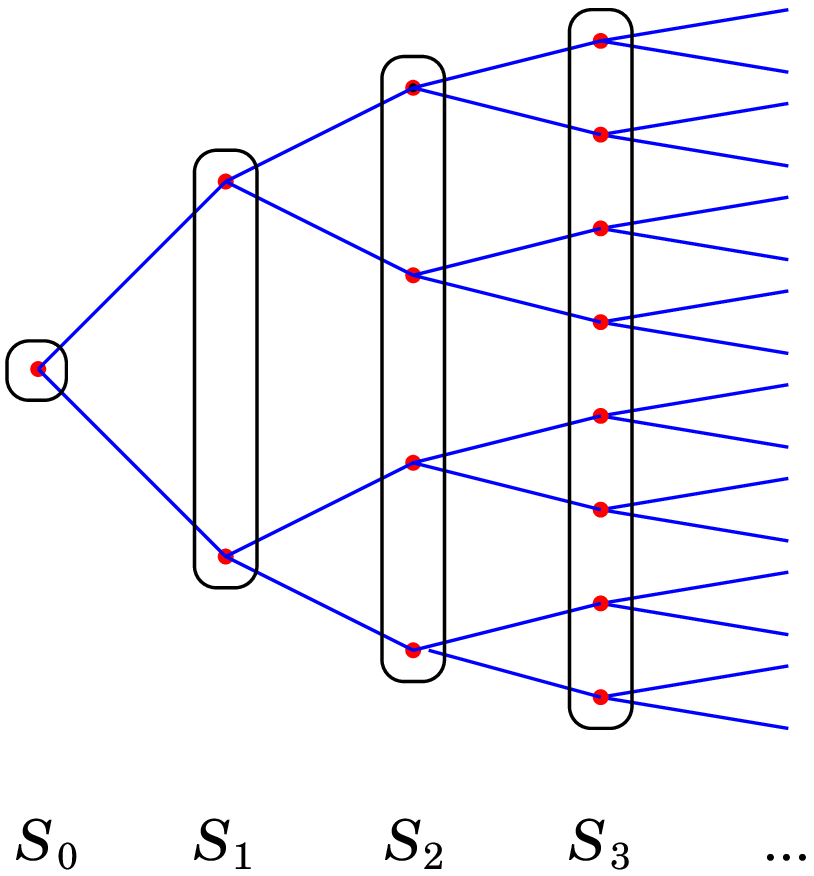}}
\centerline{\sl Spheres in a binary tree}
\end{figure}

We will write $v \ra w$ if $v$ and $w$ are connected by an edge 
and $|w| = |v| + 1$.\\
\noindent Define
$$(\Pi\phi)(v) = \ds{\sum_{w:w\ra v}} \phi(w)$$
\noindent The adjoint of $\Pi$ can be computed by calculating
$$
\lan\psi, \Pi\phi \ran = \ds{\sum_{v} \sum_{w:w\ra v}} 
\overline{\psi}(v)\phi(w)
$$
This can be interpreted as a sum over all edges joining neighboring
spheres, where $\overline{\psi}$ and $\phi$ are evaluated at the right 
and left endpoint of the edge respectively. We have chosen to label the 
edges by their right endpoint $v$, and the sum over $w:w \ra v$ accounts 
for the possibility of several edges having the same right endpoint. If 
we choose to label the edges by their left 
endpoints instead we find that the same sum can be written as
$$
\lan\psi, \Pi\phi \ran = \ds{\sum_{v} \sum_{w:v\ra w}} 
\overline{\psi}(w)\phi(v)
$$
\noindent This shows that
$$
(\Pi^{\ast} \phi)(v) = \ds{\sum_{w:v\ra w}} \phi(w)
$$
\noindent Notice that $\Pi^{\ast}\Pi$ and $\Pi\Pi^{\ast}$ leave each 
$\ell^{2}(S_{r})$ invariant.
The action of $\Pi^{\ast}\Pi$ is given by
     \begin{equation}\label{eq1}
        (\Pi^{\ast}\Pi\phi)(v) = \ds{\sum_{w}} \phi(w) 
     \end{equation}
where the sum is extended over $w \in S_{|v|}$ that are joined to $v$ by a 
path in the graph of length two going from $v$ to some element in
$S_{|v|+1}$ and then back to $S_{|v|}$. The formula for $\Pi\Pi^{\ast}$ 
is analogous, except that the path goes in the other direction to 
$S_{|v|-1}$ and back.

We will denote by $R$ the operator of multiplication by $|v|$. When 
restricted to $\ell^{2}(S_{r})$, the operator $R$ is multiplication by $r$. 
An easy calculation shows that
   \begin{equation}\label{eq2}
     [R,\Pi] = \Pi 
   \end{equation}

We may write $L$ in terms of $\Pi$ and $\Pi^{\ast}$ by breaking the sum in 
the definition of $L$ into three pieces. We obtain 
 $$ L = \Pi + \Pi^{\ast} + L_{S} $$
where the spherical Laplacian $L_{S}$ is defined by
 $$
 (L_{S}\phi)(v) = \ds{\sum_{w:\substack{w-v \\ |w|=|v|}}} \phi(w). 
 $$

\vspace*{5mm}

\noindent {\bf Diagonalization of $L$ and definition of $A$ for a Binary 
Tree}

\bsk

In this section we will exhibit a diagonalization of the off-diagonal 
Laplacian $L$ on a binary tree.

Choose the origin to be the base of the tree and introduce polar 
co-ordinates. Since there are no edges that connect vertices within each 
sphere, $L_{S} = 0$,
and
$$ 
L = \Pi + \Pi^{\ast}.
$$

We now construct invariant subspaces $M_{n}$ for $\Pi$. Let
$Q_{0,0} = \ell^{2}(S_{0})$ and define $Q_{0,r} = \Pi^{r}Q_{0,0}$.
Let
$$
M_0 = \ds{\bigoplus_{r=0}^{\infty}} \ Q_{0,r}.
$$

To define $Q_{n,r}$ and $M_{n}$ for $n > 0$ we proceed recursively. 
Suppose that $Q_{m,s}$ have been defined whenever $m < n$ and $s \geq m$. 
Let $Q_{n,n}$ be the orthogonal complement in $\ell^{2}(S_{n})$ of all 
previously defined subspaces,
$$ 
Q_{n,n} = \ell^{2}(S_{n})\ominus(Q_{0,n}\oplus \cdots \oplus Q_{n-1,n}).
$$
For $r=n+j$, $j > 0$, define $Q_{n,r} = \Pi^{j}Q_{n,n}$, and
$$
M_{n} = {\ds \bigoplus_{r=n}^{\infty} \ Q_{n,r} = \bigoplus_{j=0}^{\infty}} 
\ Q_{n,n+j}.
$$
Schematically this gives:\pagebreak

\def\h1{\hspace*{1cm}}
\def\hs{\hspace*{1.3cm}}
\def\hh{\hspace*{0.5cm}}

\vspace*{0.5cm}
\hspace*{3.33cm} $\rule[-6mm]{0.1mm}{11mm} \h1 \ell^2(S_0) \h1 \ell^2(S_1) \h1 
\ell^2(S_2) \h1 \ell^2(S_3) \h1$ \\ 
\hspace*{3cm}$\rule[0.5mm]{9.5cm}{0.1mm}$ \\
\hspace*{3cm}$M_{0} \hh \rule[-2mm]{0.1mm}{12mm} \hs Q_{0,0} \hs Q_{0,1} \hs Q_{0,2} 
\hs Q_{0,3} \hs \ldots$ \\
\hspace*{3cm}$M_{1} \hh \rule[-2mm]{0.1mm}{12mm} \hspace*{3.3cm}  Q_{1,1} \hs 
Q_{1,2} \hs Q_{1,3} \hs \ldots $ \\
\hspace*{3cm}$M_{2} \hh \rule[-2mm]{0.1mm}{12mm} \hspace*{5.3cm}  Q_{2,2} \hs 
Q_{2,3} \hs \ldots$ \\
$\hspace*{3.3cm} \vdots  \hspace*{8.3cm} \vdots $
\begin{center}
{\sl Orthogonal Subspace decomposition}
\end{center}

\begin{lemma}\label{l1} The Hilbert space $\ell^{2}(V)$ can be written as an 
orthogonal direct sum
$$ 
\ell^{2}(V) = \ds{\bigoplus_{n=0}^{\infty}} 
M_{n} = \ds{\bigoplus_{n=0}^{\infty} \bigoplus_{r=n}^{\infty}} \ Q_{n,r}.$$
The subspaces $M_{n}$ are invariant for $L$
\end{lemma}

\noindent {\it Proof:} Since $\Pi$ maps $\ell^{2}(S_{r})$ to 
$\ell^{2}(S_{r+1})$ it follows that each $Q_{n,r}$ is contained in
$\ell^{2}(S_{r})$. Thus, if $r \neq s$, then $Q_{n,r}$ and
$Q_{m,s}$ are orthogonal for all $n$ and $m$. 

For a binary tree, it follows 
from (\ref{eq1}) that \begin{equation}\label{eq3} 
    \Pi^{\ast}\Pi = 2I 
 \end{equation}
This implies that if $\phi$ and $\psi$ are orthogonal, then 
$\Pi\phi$ and $\Pi\psi$ are orthogonal too. Since $Q_{n,n}$
is orthogonal to $Q_{m,n}$, for $m < n$ by construction, it follows that 
$Q_{n,r}$ and $Q_{m,r}$, for $r \geq n$ are orthogonal too.
By construction $ \bigoplus_{l=0}^{r} Q_{l,r} = \ell^{2}(S_{r})$, so it 
is clear that the subspaces add up to $\ell^{2}(V)$.

By construction, each $M_{n}$ is invariant for $\Pi$. That they are 
invariant for $\Pi^{\ast}$ follows from (\ref{eq3}) as follows. It suffices 
to show that each $Q_{n,r}$ is mapped to $M_{n}$ under $\Pi^{\ast}$. 
Suppose that $\phi \in Q_{n,r}$ for $r=n+j$, $j \geq 1$. Then 
$\phi = \Pi^{j}\chi$ for 
$\chi \in Q_{n,n}$. Hence $\Pi^{\ast}\phi = 2\Pi^{j-1}\chi \in Q_{n,r-1}$. 
On the other hand, if $\phi \in Q_{n,n}$, then 
$\Pi^{\ast}\phi \in \ell^{2}(S_{n-1})$. Suppose that $\psi \in Q_{l,n-1}$ for 
some $l \leq n-1$. Since $\Pi\psi \in Q_{l,n}$ for some $l \leq n-1$ and
$\phi \in Q_{n,n}$, we have $\lan \psi, \Pi^{\ast}\phi \ran = \lan \Pi\psi, 
\phi \ran = 0$. 
Hence $\Pi^{\ast}\phi$ is orthogonal to each $Q_{l,n-1}$, which implies that 
$\Pi^{\ast}\phi =0$.
Thus each $M_{n}$ is invariant for $\Pi$ and $\Pi^{\ast}$, and hence 
for $L$. $\Box$

Since each $M_{n}$ is an invariant subspace for $L$ we can decompose 
$L = \oplus_{n=0}^{\infty} L_{n}$, where $L_{n}$ is the restriction of $L$ 
to $M_{n}$. We now diagonalize $L_{n}$.

We begin by writing a vector $\phi$ in $M_{n}$ as
$$
\phi = \oplus_{j=0}^{\infty} \phi_{n+j}
$$
where $\phi_{n+j} \in Q_{n,n+j}$. We want to obtain an isomorphism between
$M_{n}$ and $l^{2}(\Z^{+},Q_{n,n})$ the space of $Q_{n,n}$ valued sequences. 

We first note that $(\frac{1}{\sqrt{2}}\Pi)^{j}$ (not $\Pi^{j}$) defines an 
isometry between $Q_{n,n}$ and $Q_{n,n+j}$ for all $j$ and
that any $\phi \in M_{n}$ can be written as
\begin{equation}\label{eq4}
  \phi = \oplus_{j=0}^{\infty} \left(\ds{\frac{1}{\sqrt{2}}}\Pi \right)^{j} \ 
  \chi_{n+j}
\end{equation}
 for a sequence of vectors $\chi_{n}, \chi_{n+1}, \ldots \in Q_{n,n}$. 

Under this representation $M_{n}$ and $l^{2}(\Z^{+},Q_{n,n})$ are isomorphic,
since $$\lan \phi, \phi \ran = \sum_{j=0}^{\infty}
\lan \phi_{n+j}, \phi_{n+j} \ran = \sum_{j=0}^{\infty} \lan \chi_{n+j}, 
\chi_{n+j} \ran = \lan W\phi, W\phi \ran$$ by the above isometry where 
$W$ denotes the isomorphism.

In this representation, the operator $\frac{1}{\sqrt{2}}\Pi$ acts as a shift 
to the right, while $\frac{1}{\sqrt{2}}\Pi^{\ast}$ is a shift to the left
with kernel $Q_{n,n}$.

Now let
$$ 
U:M_{n} \cong \ell^{2}(\Z^{+},Q_{n,n}) \ra 
                                 L^{2}_{\text{odd}}([-\pi, \pi],d\theta)
$$
denote the unitary map defined by
$$
U((\alpha_{n},\alpha_{n+1},\ldots)) = 
\frac{1}{\sqrt{\pi}} \ds{\sum_{j=0}^{\infty}} \ \alpha_{n+j}\sin((j+1)\theta)
$$

\begin{lemma}\label{l2} 
$$UL_{n}U^{\ast} = 2\sqrt{2}\cos(\theta) $$
\end{lemma}
\noindent {\it Proof:} The proof is a straightforward calculation. $\Box$

This lemma shows that the spectrum of $L_{n}$ is $[-2\sqrt{2}, 2\sqrt{2}]$, 
and is absolutely continuous. Thus the spectrum of $L$ is also  
$[-2\sqrt{2}, 2\sqrt{2}]$ with infinite multiplicity.

This representation motivates the choice of a conjugate operator. For a 
general multiplication operator $\omega(\theta)$, a natural conjugate operator
is $A_{\omega} = \frac{i}{2}(\omega'\frac{d}{d\theta} + \frac{d}{d\theta}\omega')$, 
since
$$
[\omega, iA_{\omega}] = {\omega'}^{2}
$$
which is positive away from the critical points of $\omega$. 

In the present case the natural conjugate operator is therefore
$$
UA_{n}U^{\ast} = -i\sqrt{2}\left( \sin(\theta)\ds{\frac{d}{d\theta}} 
                         + \ds{\frac{d}{d\theta}}\sin(\theta) 
                   \right)
$$
and a calculation now shows that on $M_{n}$

$$
iA_{n} = U^{\ast} \sqrt{2}\left( \sin(\theta) {\ds \frac{d}{d\theta} 
                         + \frac{d}{d\theta} } \sin(\theta)   \right) U = 
                 (R - n + \ds{\frac{1}{2}})\Pi - 
                         \Pi^{\ast}(R - n + \ds{\frac{1}{2}}) 
 $$\\
Therefore a natural conjugate operator for $L$ is 
$\oplus_{n=0}^{\infty} A_{n}$.
If we let $P_{n}$ denote the projection onto $M_{n}$, and define  
$$
N = \ds{\sum_{n=1}^{\infty}} n P_{n},
$$
the conjugate operator for $L$ can be written as
$$
iA =  (R-N+\ds{\frac{1}{2}})\Pi - \Pi^{\ast}(R-N + \ds{\frac{1}{2}})
$$

\bsk

\noindent {\bf Matrix elements of $A$}

\bsk

We will need estimates on the matrix elements of $A$. Let $\delta_{w}$ denote 
the standard basis element in $\ell^{2}(V)$ defined by 
$\delta_{w}(v) = \delta_{w,v}$. We wish to estimate the matrix elements 
$\lan \delta_{v}, iA\delta_{w} \ran$.

Using the formula
$$
\Pi \delta_{w} = \ds{\sum_{z:w\ra z}} \delta_{z}
$$
we find that

 \begin{equation}\label{eq5}
   \begin{array}{ccl}
     \lan \delta_{v}, iA\delta_{w} \ran & = &  
     \lan \delta_{v}, \left( (R-N+\ds{\frac{1}{2}})\Pi - 
                              \Pi^{\ast}(R -N + \ds{\frac{1}{2}})
                      \right)
       \delta_{w} \ran \\ 
        & = & \left\{ \begin{array}{ccc} 
                 (|v| + \frac{1}{2}) \delta(w\ra v) - 
                      {\ds \sum_{z:w\ra z}} \lan \delta_{v},N\delta_{z} \ran 
                          & \mbox{if} & |w| = |v| - 1 \\
                -(|w| + \frac{1}{2}) \delta(v\ra w) + 
                     {\ds \sum_{z:v\ra z}} \lan \delta_{z}, N\delta_{w} \ran 
                          & \mbox{if} & |w| = |v| + 1 \\
                      0   &           & \mbox{otherwise}
                      \end{array} 
              \right.
   \end{array}
 \end{equation}
\noindent Here $\delta(w \ra v)$ is equal to 1 if $w \ra v$ and 0 otherwise.

To estimate the matrix elements of $N$ appearing in this formula, we introduce
at this point an explicit basis for each $Q_{n,r}$. When $n=0$ we have that 
$Q_{0,r} = \Pi^{r}\ell^{2}(S_{0})$ is one dimensional and consists of all 
vectors $\phi(v)$ in $\ell^{2}(S_{r})$ such that $\phi(v)$ has the same value 
for all $v$. An orthonormal basis for $Q_{0,r}$ is therefore 
the single vector
$$ 
\rho_{0,r,0} = 2^{-r/2}[1,1,\ldots,1]
$$
The space $Q_{1,1}$ is the orthogonal complement in $\ell^{2}(S_{1})$ of 
$Q_{0,1}$. Thus $Q_{1,1}$ is also one dimensional and has orthonormal basis
$$ 
\rho_{1,1,0} = 2^{-1/2}[1,-1]
$$
Pushing the vector forward along the tree using $\Pi$ and normalizing gives
$$ 
\rho_{1,r,0} = 2^{-r/2}[1,1,\ldots,1,-1,-1,\ldots,-1]
$$
as a basis for $Q_{1,r}$ where half the entries are 1 and the other half -1.

The space $Q_{2,2}$ is the orthogonal complement in $\ell^{2}(S_{2})$ of 
$Q_{0,1} \oplus Q_{1,1}$. Since dim$(\ell^{2}(S_{2})) = 4$ the space 
$Q_{2,2}$ is two dimensional, it has orthogonal basis
$$ 
\rho_{2,2,0} = 2^{-1/2}[1,-1,0,0] \ \ \ \ \rho_{2,2,1} = 2^{-1/2}[0,0,1,-1]
$$
Pushing these vectors forward along the tree using $\Pi$ and normalizing yields
$\rho_{2,r,0}$ and $\rho_{2,r,1}$.

Continuing in this fashion we define the orthogonal basis $\rho_{n,r,k}$ with 
$k=0,\ldots,2^{\text{max}\{n-1,0\} }-1$. When we fix the second index $r$, the 
vectors $\rho_{n,r,k}$ are the Haar basis for $\ell^{2}(S_{r})$.
Upon defining a partial order on the Haar basis elements for 
$\ell^{2}(S_{r})$ using inclusion of supports, the Haar basis functions, as
illustrated on the next page for $r=4$, naturally form a binary tree with 
$r$ levels, extended by an extra vertex at its base.
\begin{figure}[h]
   \centerline{\epsfysize=8.5cm
   \epsfbox{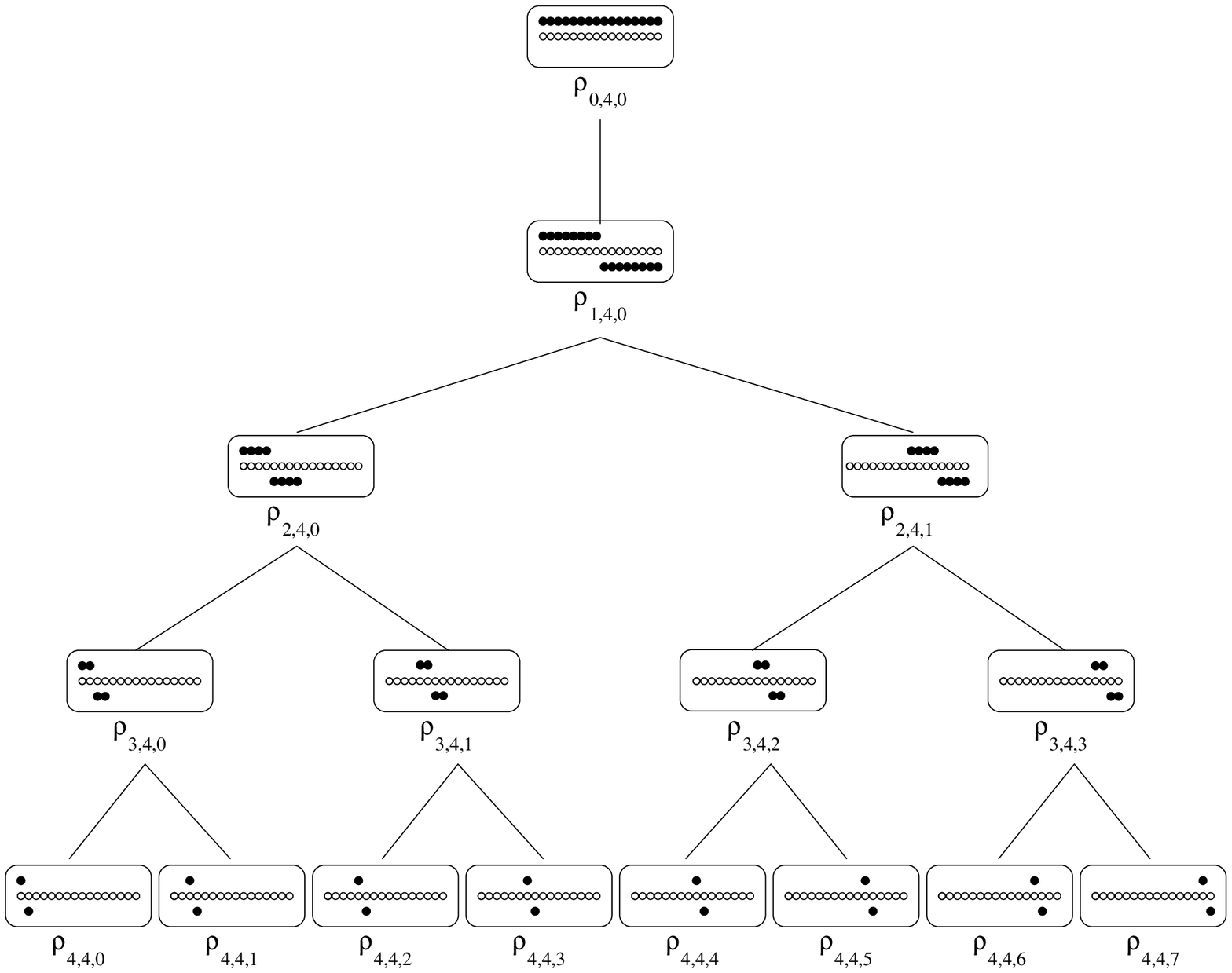}}
\medskip
\centerline{\sl The Haar basis}
\end{figure}
\begin{lemma}\label{l3} Let $z$,$w \in S_{r}$. If $z \neq w$, let $N(z,w)$ 
denote the largest value of $n$ for which both $z$ and $w$ lie in the support 
of a single basis function. Then
$$
\langle \delta_{z}, N\delta_{w} \rangle = 
        \left\{ \begin{array}{ccc} 
               r - 1 + 2^{-r}        & \mbox{if} & z=w \\
              -2^{N(z,w)-r} + 2^{-r} & \mbox{if} & z \neq w 
                \end{array} 
        \right.
$$
\end{lemma}

\noindent {\it Proof: } Fix $r$ and label the elements of the Haar basis by 
the vertices $\alpha$ in the associated (extended) basis 
binary tree. Then
$$
N\delta_{w} = \ds{\sum_{\alpha}} n(\alpha) \rho_{\alpha}(w) 
              \rho_{\alpha} 
$$
where $n(\alpha)$ denotes the level of $\alpha$ in the tree and the sum is 
taken over the one $\alpha$ at each level for which $\rho_{\alpha}(w) \neq 0$ 

Suppose $z = w$. Then, for the one $\alpha$ at level $n(\alpha) = n$,
for which $\rho_{\alpha}(w) \neq 0$, we have 
$\rho_{\alpha}(w)^{2} = 2^{-r+n-1}$. Thus
$$
\langle \delta_{w}, N\delta_{w} \rangle = {\ds \sum_{n=1}^{r}} n2^{-r+n-1} = 
                                          r - 1 + 2^{-r}
$$

Now suppose $z \neq w$. Then for the one
$\alpha$ at level $n$ for which $\rho_{\alpha}(w) \neq 0$ we have
$$
\rho_{\alpha}(z) = \left\{ 
                   \begin{array}{cc}
                      - \rho_{\alpha}(w) & \mbox{ if } n = N(z,w) \\ 
                        \rho_{\alpha}(w) & \mbox{ if } n < N(z,w) \\
                        0 & n > N(z,w) \\
                   \end{array} \right.
$$
Hence 
$$
\begin{array}{ccl} \langle \delta_{z}, N\delta_{w} \rangle 
  & = & {\ds \sum_{n=1}^{N(z,w)-1}} n2^{-r+n-1} - N(z,w)2^{-r+N(z,w)-1} \\
  & = & -2^{-r + N(z,w)} + 2^{-r} \ph
\end{array}
$$
\noindent $\Box$
                
\begin{lemma}\label{l4}
$$ \ds{\sum_{w}} |\langle \delta_{v},iA\delta_{w} \rangle | = O(|v|) $$
\end{lemma}

\noindent {\it Proof:} Using (\ref{eq5}) we find

\begin{equation}\label{eq6}
\begin{array}{rl}
{\ds \sum_w} |\lan \delta_v,iA\delta_w \ran| & 
\leq \ds{\sum_{w:|w|=|v| - 1}} \left( |v|+ \frac{1}{2} \right) \delta(w \ra v) 
  + {\ds \sum_{w:|w|=|v| - 1} \sum_{z:w \ra z}} 
            |\lan \delta_v, N \delta_z \ran | \\
 & \ \ + {\ds \sum_{w:|w|=|v|+1} \left( |w|+ \frac{1}{2} \right)} 
       \delta(v \ra w) 
  + {\ds \sum_{w:|w|=|v|+1}\sum_{z:v \ra z}} |\lan \delta_z, N \delta_w \ran | 
\end{array}
\end{equation}
Since there is only one $w$ with $w \ra v$ we have
$$
\sum_{|w|=|v| - 1} \left( |v| + \frac{1}{2} \right) \delta( w \ra v ) = |v| + \frac{1}{2}.
$$
Since there are exactly two $w$ with $v \ra w$,
$$
\sum_{|w|=|v|+1} \left(|w|+ \frac{1}{2} \right) \delta( v \ra w)= 2(|v|+1)+1.
$$
To estimate the remaining two terms in (\ref{eq6}), we begin with
\begin{equation}
\begin{array}{rcl}
\ds{ \sum_{z:|z|=|v|}} |\lan \delta_v, N \delta_z \ran | 
    & = & |\lan \delta_v, N\delta_v \ran| + 
        \ds{ \sum_{z:{{ |z| = |v|}\atop{z \neq v}}}} 
                |\lan \delta_v,N\delta_z \ran | \\
    & = & |v|-1+2^{-|v|} + 
         {\ds \sum_{z:{{ |z| = |v| }\atop{ z \neq v}}}}2^{N(v,z)-|v|} - 2^{-|v|} \\
   & \leq & |v| + {\ds \sum_{z:{{ |z| = |v| }\atop{ z \neq v}}}}2^{N(v,z)-|v|} 
\end{array}
\end{equation}
Since there are $2^{|v|-N(v,z)}$ $z$'s associated to each value of $N(v,z)=n$ 

$$
\begin{array}{rcl}
\ds{ \sum_{z:|z|=|v|}} |\lan \delta_v, N \delta_z \ran | 
 & \leq & |v| \ + \ {\ds \sum_{n=1}^{|v|}2^{n-|v|}
                 \cdot\sum_{z:|z|=|v|, z \neq v, N(v,z)=n}} 1 \\ 
 &  =   & |v| \ + \ {\ds \sum_{n=1}^{|v|}}2^{n-|v|}2^{|v|-n} \\
 &  =   & 2|v|
\end{array}
$$
Therefore
$$
\sum_{w:|w|=|v| - 1}\sum_{z:w\ra z}|\lan \delta_v, N\delta_z \ran | \ = 
\sum_{z:|z|=|v|}| \lan \delta_v, N\delta_z \ran|
\leq 2|v|
$$
and
$$
\sum_{w:|w|=|v|+1}\sum_{z:v\ra z}|\lan \delta_z, N\delta_w \ran |= \sum_{z:v\ra z}\sum_{w:|w|=|z|}
|\lan \delta_z,N\delta_w \ran| \leq \sum_{z:v\ra z}2|v|=4|v|.
$$
Thus each term in (\ref{eq6}) is $O(|v|)$ and the proof is complete.
$\Box$

\bsk

\noindent {\bf The Mourre estimate}

\bsk

We begin with the commutator formula for $L$. This is just a disguised form
of the formula
$$
[2\sqrt{2}\cos(\theta), \sqrt{2}(\sin(\theta)\frac{d}{d\theta} 
            + \frac{d}{d\theta}\sin(\theta))] = 8\sin^{2}(\theta).
$$

\begin{lemma}\label{l5}
$$ [L,iA] = 8 - L^{2} $$
\end{lemma}
\noindent {\it Proof:} Since $\Pi$ and $\Pi^{\ast}$ commute with $N$, we have
$$
\begin{array}{rcl}
      [L,iA] & = & [\Pi + \Pi^{\ast}, (R - N + \frac{1}{2}) \Pi] + 
                      \mbox{ adjoint}\\
             & = & [\Pi,(R - N + \frac{1}{2})] \Pi + 
                   [\Pi^{\ast}, (R - N + \frac{1}{2})] \Pi + 
                   (R-N + \frac{1}{2})[\Pi^{\ast}, \Pi] + \mbox{ adjoint} \\
             & = & [\Pi, R]\Pi + [\Pi^{\ast}, R]\Pi + 
                   (R-N + \frac{1}{2})[\Pi^{\ast}, \Pi] + \mbox{ adjoint} \\
             & = & -\Pi^{2} + \Pi^{\ast}\Pi + (R-N + 
                   \frac{1}{2})[\Pi^{\ast}, \Pi] + \mbox{ adjoint} 
\end{array}
$$
where adjoint applies to all the previous terms.
\noindent Here we used (\ref{eq2}). Now notice that $[\Pi^{\ast}, \Pi]$ is a 
projection onto the sum of the initial subspaces $Q_{n,n}$ and that is 
precisely the kernel of the operator $R-N$ since on this sum $R=N$. Thus 
$(R-N)[\Pi^{\ast}, \Pi] = 0$ and, 
$$
\begin{array}{rcl}
     [L,iA] & = & -\Pi^{2} + \Pi^{\ast}\Pi + 
                   \frac{1}{2}[\Pi^{\ast},\Pi] + \mbox{ adjoint } \\
            & = & -\Pi^{2} - (\Pi^{\ast})^{2} + 3\Pi^{\ast}\Pi - 
                            \Pi\Pi^{\ast} \\
            & = & 4\Pi^{\ast}\Pi - (\Pi + \Pi^{\ast})^{2} \\
            & = & 8 - L^{2}
       \end{array}$$
\noindent $\Box$
\begin{lemma}\label{l6} Suppose that
$$\sup_{w:|w| = |v| \pm 1} |q(v) - q(w)| = o(|v|^{-1}),$$
as $|v| \ra \infty$, then $[q,iA]$ is compact.
\end{lemma}

\noindent {\it Proof:} Let $\Lambda_{n}$ denote the projection onto 
$\oplus_{r=0}^{n} l^{2}(S_{r})$. We will show that 
$||[q,iA] -[q,iA]\Lambda_{n}||= ||[q,iA](1 - \Lambda_{n})|| \ra 0$ as
$n \ra \infty$. This shows that $[q,iA]$ is approximated in norm by the 
finite rank operator $[q,iA]\Lambda_{n}$, and hence compact. 

The matrix elements of $[q,iA](1-\Lambda_{n})$ are given by
$$ 
\lan \delta_{v}, [q,iA](1-\Lambda_{n}) \delta_{w} \ran = (q(v) - q(w)) 
                                      \lan \delta_{v}, iA\delta_{w} \ran
$$
provided $|w| > n$, and 0 if $|w| \leq n$. Using Schur's lemma 
(the $\ell^{1}-\ell^{\infty}$ bound), the fact that the matrix element of 
$\lan \delta_{v}, iA\delta_{w} \ran$ are non-zero only for $|w| = |v| \pm 1$, 
the decay hypothesis on $q$, and Lemma \ref{l4} we find that
$$
\begin{array}{rcl}
|| [q,iA](1-\Lambda_{n})|| & \leq & {\ds \sup_{v} \sum_{w:|w| > n}} 
|q(v) - q(w)| | \lan \delta_{v},i A\delta_{w} \ran | \\
& \leq & {\ds \sup_{v:|v| > n-1}} o(|v|^{-1}) {\ds \sum_{w}}  
| \lan \delta_{v}, i A\delta_{w} \ran | \\
& \leq & {\ds \sup_{v:|v| > n-1}} o(|v|^{-1})O(|v|)
\end{array}
$$
which tends to zero for large $n$.
$\Box$

\begin{lemma}\label{l7} Suppose that
$$
\sup_{\substack{w:|w|=|v| \pm 1 \\ z:|z| = |v| \pm 2}} 
           |q(v) + q(z) - 2q(w)| = O(|v|^{-2}),
$$ 
as $|v| \ra \infty$, then $[[q,iA],iA]$ is bounded.
\end{lemma}

\noindent {\it Proof:} The matrix elements of $[[q,iA],iA]$ are given by
$$
\begin{array}{rcl} 
   \lan \delta_{v}, [[q,iA],iA]\delta_{z} \ran 
 & = & \lan \delta_{v}, [q,iA]iA - iA[q,iA]\delta_{z} \ran \\
 & = & {\ds \sum_{w} } \lan \delta_{v}, [q,iA]\delta_{w} \ran 
       \lan \delta_{w}, iA\delta_{z} \ran - \lan \delta_{v}, iA \delta_{w} \ran
       \lan \delta_{w}, [q,iA]\delta_{z} \ran \\
 & = &  {\ds \sum_{w} } (q(v) + q(z) - 2q(w))\lan \delta_{v}, iA\delta_{w} \ran
        \lan \delta_{w}, iA\delta_{z} \ran
\end{array}.
$$
Thus as in Lemma \ref{l6},
$$
\begin{array}{rcl}
   ||[q,iA],iA]|| & \leq & {\ds \sup_{v}} \ds{\sum_{z} \sum_{w}} 
             |q(v) + q(z) - 2q(w)||\lan \delta_{v}, iA\delta_{w} \ran| 
                                  |\lan \delta_{w}, iA\delta_{z} \ran| \\
                  & \leq & {\ds \sup_{v}} \ O(|v|^{-2}) \ds{\sum_{w}} 
             |\lan \delta_{v}, iA\delta_{w} \ran| \ds{\sum_{z}} 
                                  |\lan \delta_{w}, iA\delta_{z} \ran| \\
                  &  =   & {\ds \sup_{v}} \ O(|v|^{-2})O(|v|)O(|v|) \leq C
\end{array}
$$
$\Box$

\begin{lemma}\label{l8} Suppose that $q(v) \ra 0$ as $|v| \ra \infty$. Let 
$E$ denote a smoothed out spectral projection. Then $E(L) - E(L + q)$ is 
compact.
\end{lemma}

\noindent {\it Proof:}
This follows from the compactness of 
$(L-z)^{-1} - (L + q -z)^{-1} = (L - z)^{-1}q(L + q -z)^{-1}$
and a Stone-Weierstrass approximation argument (see \cite{CFKS}). $\Box$

Now we can prove the Mourre estimate for $L + q$ and $A$.

\begin{theorem}\label{t9} Suppose that $q(v) \ra 0$ as $|v| \ra \infty$. Assume 
that
$$
\sup_{w:|w| = |v| \pm 1} |q(v) - q(w)| = o(|v|^{-1}).
$$
as $|v| \ra \infty$. Let $E$ denote a smoothed out spectral projection whose 
support is properly contained in the interval $(-2\sqrt{2},2\sqrt{2})$. Then 
there exists a compact operator $K$ and a positive number $\alpha$ such that
$$
E(L + q)[L+q,iA]E(L+q) \geq \alpha E^{2}(L+q) + K.
$$
\end{theorem}

\noindent {\it Proof:} By compactness of $[q,iA]$ and $E(L+q) - E(L)$ we have
$$
\begin{array}{rcl} E(L + q)[L+q,iA]E(L+q) 
                      & = & E(L)[L,iA]E(L) + K\\
                      & = & E(L)(8 - L^{2})E(L) + K 
\end{array}
$$

\noindent On the support of $E(L)$, $8 -L^{2} \geq \alpha$ for some positive 
$\alpha$, which gives
         $$E(L + q)[L+q,iA]E(L+q) \geq \alpha E^{2}(L+q) + K$$
where the compact term $E^{2}(L) - E^{2}(L+q)$ has been added into $K$, 
this completes the proof.
\noindent $\Box$

\end{document}